\documentclass[12pt,dvipdfmx]{emulateapj}

\usepackage{graphicx} 
\usepackage{color}

\shorttitle{3C~84 shell}
\shortauthors{M. Kino et al.}

\begin{document}

\title{Fossil shell in 3C~84 as TeV $\gamma$-ray emitter and cosmic-ray accelerator}
\author{
M. Kino\altaffilmark{1,2},
H. Ito\altaffilmark{3}, 
K. Wajima\altaffilmark{1},
N. Kawakatu\altaffilmark{4},
H. Nagai\altaffilmark{2},
R. Itoh\altaffilmark{5}}

\altaffiltext{1}{Korea Astronomy and Space Science Institute, 
776 Daedeokdae-ro, Yuseong-gu, 
Daejeon 34055, Republic of Korea} 
\email{kino@kasi.re.kr}
\altaffiltext{2}{National Astronomical Observatory of Japan
                 2-21-1 Osawa, Mitaka, Tokyo, 181-8588, Japan}
\altaffiltext{3}{Astrophysical Big Bang Laboratory, RIKEN, Saitama 351-0198, Japan}
\altaffiltext{4}{National Institute of Technology, Kure College, 2-2-11 Agaminami, Kure, Hiroshima, 737-8506, Japan}
\altaffiltext{5}{Department of Physics, Tokyo Institute of Technology, 2-12-1 Ohokayama, Meguro,
Tokyo 152-8551, Japan}



\begin{abstract}

We explore physical properties of
the shocked external medium (i.e., a shell)
in 3C~84
associated with the recurrent radio lobe 
born  around 1960.
In the previous work of Ito et al.,
we investigated a dynamical and radiative evolution 
of such a shell after the central engine stops the jet launching
and we found that a fossil shell emission overwhelms that of 
the rapidly fading radio lobe.
We apply this model to 3C~84 and find the followings:
(i) The fossil shell made of shocked diffuse ambient matter with the number density 
of $0.3~{\rm cm^{-3}}$ radiates bright Inverse-Compton (IC) emission
with the seed photons of  the radio emission from the central compact region
and the IC emission is above the sensitivity threshold of the Cherenkov Telescope Array (CTA).
(ii) 
When the fossil shell is produced in a geometrically　thick 
ionized plasma with the number density of $10^{3}~{\rm cm^{-3}}$
and 
the field strength in the shell may reach  about $17$~mG 
in the presence of magnetic fields amplification 
and  the radio emission becomes comparable to the sensitivity 
of deep imaging VLBI observations.
A possible production of ultra high energy cosmic-rays (UHECRs) 
in the dense shocked plasma is also argued.

\end{abstract}

\keywords{
galaxies: active 
radio continuum: galaxies
gamma rays: galaxies 
galaxies: individual (3C 84, NGC 1275)}

\section{Introduction}
\label{sec:intro}

Radio-loud active galactic nuclei (AGNs) are among the most
powerful objects in the Universe. 
Interactions between jets and ambient medium
drive strong collisionless shocks in the 
surrounding external medium. 
Therefore,
AGN jets can provide us with
a great variety of important information
about fundamental physical processes in collisionless  shocks 
(e.g., Marcowith et al. 2016 for a review).
An AGN  jet is thought to play an important role for
radio-mode feedback against interstellar matter in its host galaxy
(e.g., Fabian 2012 for review).
According to the standard picture of jets in AGNs
(e.g., Begelman et al. 1984), a jet is enveloped in a cocoon
consisting of shocked jet material.
The hot cocoon's pressure drives
 the forward shocks and  the forward-shocked external medium produces the shell structure.
Although the shell is a fundamental ingredient,  physical properties 
of the shell have not been well studied
because they are faint radio-quiet emitters (Carilli et al. 1988)
and still undetected at radio wavelengths.

Recent theoretical work on forward shocks have 
provided us with basic properties of forward shocks in AGNs
 such as
 predicted high energy $\gamma$-ray emission
(e.g., Fujita et al. 2007; Ito et al. 2011; Kino et al. 2013; Ito et al. 2015 (hereafter I15)), 
possible cosmic-ray production and 
magnetic field amplification
(e.g., Berezhko 2008).
However, comparisons between these theoretical predictions
and observations  are poorly  studied  because of the paucity of 
information about  the surrounding external medium
in which forward shocks are driven.

The compact radio source 3C~84
(also known as NGC1275)
is one of the nearby ($z=0.018$) best-studied radio galaxies.
3C~84 shows intermittent jet activity (e.g., Nagai et al. 2010).
Once the jet activity stops, the radio emission of the lobes fades out
rapidly making the shell dominate the radio emission (I15). 
For this reason, the intermittent radio activity
observed in 3C 84 makes this source a good candidate for observational studying the shell emission.
Comparison between  model spectra 
and observed non-thermal emission
generally provides us with straightforward limits of 
magnetic fields (hereafter $B$-fields) strength  and number density, 
particle (electron) acceleration efficiency 
in the shell.

The goal of this paper is 
to explore basic properties of the fossil shell in 3C~84
and its detectability.
We will examine whether we can
constrain the magnetic field strength and particle
acceleration efficiency in the shell by comparing 
the theoretical predictions with the observations.

The layout of this paper is as follows.
In \S 2, we briefly review the model following the work of I15.
In \S 3,  physical quantities
associated with 3C~84 radio lobe and surrounding 
environment are summarized.
In \S 4, the shell emission spectra
predicted by applying the model to 3C~84
using the quantities shown in \S 3.
Summary is presented in \S 5.
In this work, we define the radio spectral index $\alpha_{R}$
as $S_{\nu}\propto \nu^{-\alpha_{R}}$.
The cosmological parameters used here are as follows;
$H_{0} = 71~{\rm km/s/Mpc}$, 
$\Omega_{\lambda} = 0.73$ and 
$\Omega_{m} = 0.27$
(e.g., Komatsu et al. 2011).
The redshift of 3C~84 ($z=0.018$) is located 
at the luminosity distance  $D_{\rm L}=75$~Mpc and 
it corresponds to 0.35~pc mas$^{-1}$.

\section{Model}

Details of the model of pressure-driven 
expanding jet-remnant  system have already been 
well established 
(Ostriker \& McKee 1988; I15 and references therein).
In this paper,  we  simply follow it in the present work.

\begin{figure} 
\includegraphics
{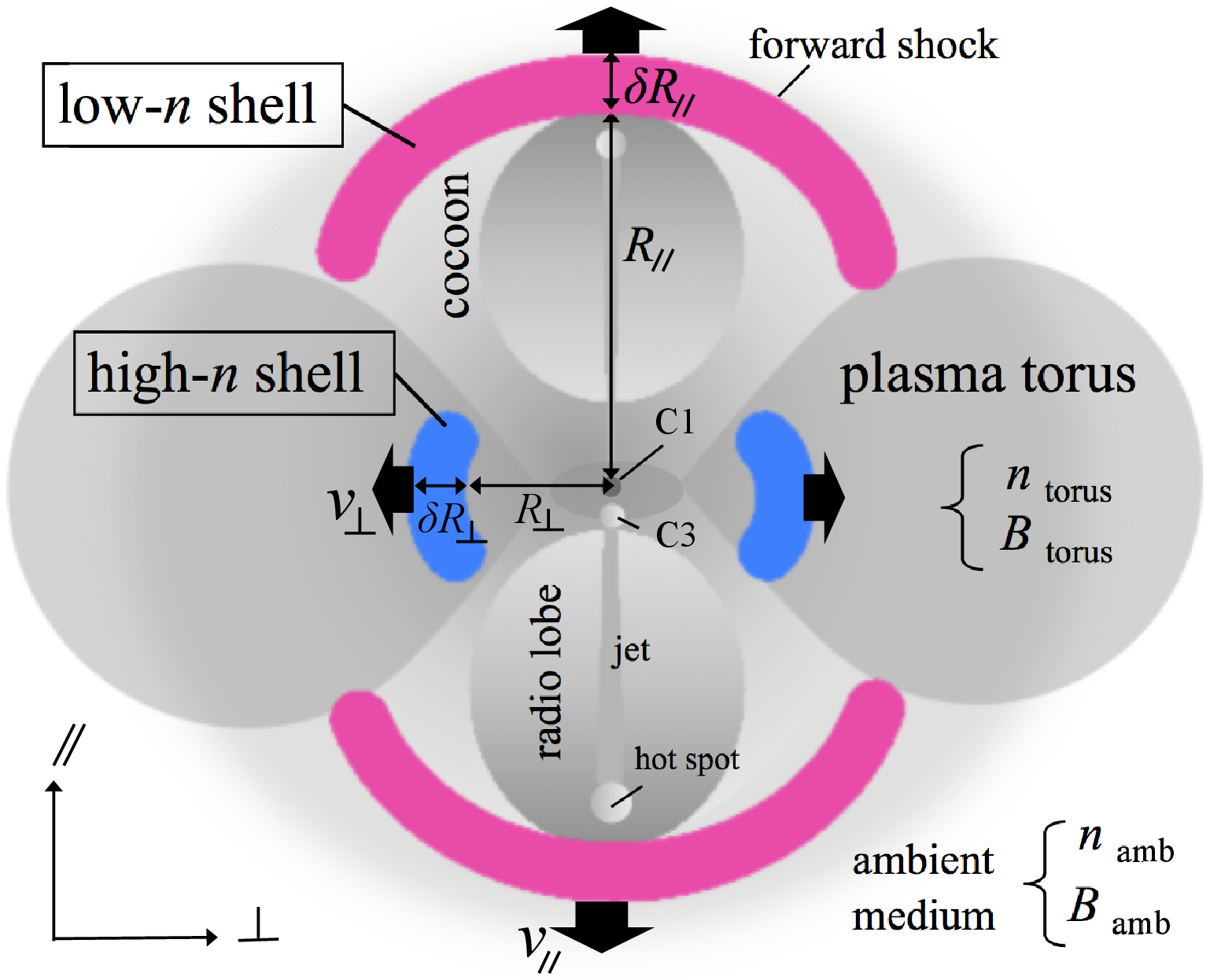}
\caption
{
A schematic picture of the jet and external medium interaction in 3C~84.
In this picture, side-on view (i.e., the jet viewing angle as 90 degree)
is adopted for convenience. }
\label{fig:cartoon}
\end{figure}

In  Figure~\ref{fig:cartoon}, we take the specific case of 3C~84.
The kinetic energy of the jets is dissipated via the termination shock 
at the hot spots and deposited
into the cocoon (radio lobes) with its radius $R$ and the shell with 
its width $\delta R$. 
In the present work, we estimate $R_{\parallel} \approx 10$~pc in 2015
(see Table ~1).
The cocoon is inflated by its internal energy.
The cocoon drives the forward shock propagating 
in the external medium and the forward-shocked region is identical to the shell.
The C3 component near the nucleus is the well-known 
newborn component (hot spot/radio lobe) which propagates southward
(Nagai et al. 2010; Suzuki et al. 2012).
O'Dea et al. (1984) and Walker et al. (2000) clarified the existence
of dense ionized gas which causes FFA of northern jet. 
In this picture, we describe the dense gas as a " plasma torus"
which is  suggested in some of radio galaxies (e.g., Kameno et al. 2001 for NGC~1052).
Since the  actual jet axis viewing angle is not 90 degrees, 
the northern part of the jet and radio lobe are hidden by 
FFA due to the surrounding matter (Walker et al. 2000).
In Figure~\ref{fig:cartoon}, the surrounding matter is described as 
the ionized plasma torus.  
The outside of the plasma torus would match a dust torus region.
Since a geometrical relation between the dust torus and the 
ionized plasma torus is highly uncertain and it is still under debate
(e.g., Netzer \& Laor 1993; 
Czerny \& Hryniewicz2011), we  do not display 
the dust torus  in Figure~\ref{fig:cartoon}.

\subsection{Dynamics}

Here we briefly summarize the dynamics of the expanding cocoon.
The cocoon radius ($R$)
is determined by the momentum balance 
between the cocoon's internal pressure and 
the ram pressure.
The mass density of surrounding external matter ($\rho_{\rm ext}$) at $R$ is defined as 
\begin{eqnarray}\label{eq:rho_ext}
\rho_{\rm ext}(R)=\rho_{0} \left(\frac{R}{R_{0}}\right)^{-\alpha}  ,
\end{eqnarray}
where $\rho_{0}$, $R_{0}$, and $\alpha$ 
are 
the reference mass density,
the reference radius, and 
the power-law index of $\rho_{\rm ext}$,
respectively.
The shell width ($\delta R$) at $R$ satisfies the relation 
$\delta R= 
(\hat{\gamma}_{\rm ext}-1)[(\hat{\gamma}_{\rm ext}+1)(3-\alpha)]^{-1} R$, 
where 
$\hat{\gamma}_{\rm ext}$
is the specific heat ratio of the external medium.

Following the previous work of I15,
we consider two phases depending on the source age ($t$):
\begin{itemize}
\item
(i) the phase in which the jet energy injection 
into the cocoon continues
($t<t_{j}$)

\item
(ii) the  phase after the jet has switched off
$(t>t_{j})$
\end{itemize}
where
$t_{j}$
denotes the duration of the jet injection.
Hereafter, we  assume that the
kinetic power of the jet ($L_{j}$) is constant in time
when $t<t_{j}$ and $L_{j}=0$ for $t\ge t_{j}$.
The bulk kinetic
energy of the jet is dissipated and deposited 
as the internal energy of the cocoon and shell. 
As for the early phase with jet energy injection into the cocoon, 
the time evolution of $R$  is given by
\begin{eqnarray}\label{eq:R}
R(t)=C R_{0}^{\frac{\alpha}{\alpha-5}}
\left(\frac{L_{j}}{\rho_{0}}\right)^{\frac{1}{5-\alpha}}
t^{\frac{3}{5-\alpha}} \quad (t<t_{j})    ,
\end{eqnarray}
where $C$ is the numerical coefficient 
and the explicit form of $C$ is shown in I15.
Note that $L_{j}/\rho_{0}$ is the key control parameter
which governs the dynamical expansion of the bubble
(Kawakatu et al. 2009a and reference therein).

After the energy injection from the jet stops, (i.e., $t>t_{j}$)  
the cocoon will rapidly lose its energy due to adiabatic expansion and
give away most of its energy into the shell within a dynamical timescale.
Hence, after $t\approx t_{j}$, the cocoon pressure becomes dynamically unimportant,
and the energy of the shell becomes dominant. 
Therefore, the behavior of the cocoon
asymptotically  follows the Sedov-Taylor expansion
and the expansion velocity is given by 
\begin{eqnarray}\label{eq:v}
v(t) = \dot{R}(t) \propto t^{-(3-\alpha)/(5-\alpha)}
\quad (t>t_{j})    .
\end{eqnarray}
During the Sedov-Taylor expansion phase, 
the adiabatic relation $P_{c} V_{c}^{\hat{\gamma}_{c}}=const$ holds
where $P_{c}$, $V_{c}$, and $\hat{\gamma}_{c}$ 
are the pressure, volume, and 
the specific heat ratio of the cocoon, respectively.
Then, we can approximately describe  
$R_{c}(t)$ 
as follows:
\begin{eqnarray}
R_{c}(t)=
R(t_{j})
\left[\frac{P_{c}(t_{j})}{P_{c}(t)}\right]^{1/3\hat{\gamma}_{c}}
 \quad (t>t_{j})     .
\end{eqnarray}
Note that the cocoon expands slower than 
the propagation speed of the forward shock and
the shell width correspondingly  becomes slightly wider 
(e.g., Figure~1 of Reynolds \& Begelman 1997).
This behavior holds also for relativistic regime and 
it is known as an expanding-coasting phase
(e.g., Piran 1999 for a review.)
The time evolution of $P$  is given by
\begin{eqnarray}\label{eq:P}
P_{c}(t)\approx P_{\rm shell}(t)=
\frac{3}{4}
\rho_{0}\dot{R}(t)^{2}
\left[\frac{R(t)}{R_{0}}\right]^{-\alpha}
 \quad (t>t_{j})   ,
\end{eqnarray}
by matching the pressures 
between the shell and cocoon at $t=t_{j}$.
For the shock jump condition between
the shell and external medium, 
the specific heat ratio of external medium 
is set as $5/3$
and it leads to $\delta R/R =1/12$.

Hereafter,  the subscripts $\parallel$ and $\perp$ 
are used for $R$ and $v$ (see Figure~\ref{fig:cartoon}).
The subscript $\parallel$  describes the direction parallel to the jet axis,
while the subscript $\perp$  corresponds to the direction perpendicular to it.

\subsection{Geometry of external medium}

Various radio observations of 3C~84 suggest the
existence of dense external ionized gas (i.e., plasma) in the central region
(e.g., O'Dea et al. 1984; Walker et al. 2000; Plambeck et al. 2014).
This circum-nuclear structure of dense ionized gas hides
 the northern radio lobe via 
free-free absorption (FFA) process (Walker et al. 2000).
Hence, shocks  driven by the cocoon propagate through both 
the diffuse ambient medium and the dense circum-nuclear structure of ionized gas.
Geometry of the dense plasma is quite uncertain.
If the dense plasma is a geometrically thin disk, 
then the volume of the shocked dense plasma may be negligibly small.
On the other hand, if the plasma shows a torus-like geometry,
the volume would not be that small.
We introduce a free parameter which  represents 
filling factors of the shocked plasma torus and
ambient matter over the shell as $f_{\rm torus}$ and 
$f_{\rm amb}$.
Since the cocoon  expands quasi-spherically in all directions,
$f_{\rm amb}+f_{\rm torus} \approx 1$ holds where
$0< f_{\rm amb} \lesssim 1$, and 
$0\lesssim f_{\rm torus}<1$.

In the present work, we examine the case when
the dense plasma has the torus-like geometry
(Figure~\ref{fig:cartoon}).
In the present work, we examine the case in which 
$f_{\rm amb} \approx f_{\rm torus} \approx 0.5$ is realized.

\subsection{Non-thermal emission}

Since the details have been already explained in
Kino et al. (2013) and I15,
here we briefly review the basic treatment of 
non-thermal electrons and photons in a shell. 
We solve the kinetic equation of 
the non-thermal electrons including the back reaction of 
radiative and adiabatic coolings.
First, 
as for the external photon field against IC process
we consider 
(1) UV photons from a standard accretion disk, 
(2) IR photons from a dust torus, 
(3) synchrotron photons from the fading radio lobe,
(4) synchrotron photons from the central compact region, and 
(5) synchrotron photons from the shell.
Second,
we include the effect of absorption via $\gamma \gamma$  interaction.
Very high-energy (VHE) photons suffer from  absorption via interaction
with various soft photons (e.g., Coppi \& Aharonian 1997). 
Here, we include the $\gamma \gamma$  absorption 
due to photons intrinsic  to the source  and 
photons from the extragalactic background light (EBL). 
The absorption opacity with respect
to the intrinsic photons can be calculated by summing
up all of the photons from (1) the shell, (2) the radio lobes,
(3) the dusty torus, and (4) the accretion disk and we multiply
the  $\gamma \gamma$ absorption factor of 
$\exp(-\tau_{\gamma \gamma})$ with the unabsorbed flux.
For simplicity, we deal with the absorption effect at the first
order and we neglect cascading effect. 
For the cosmic  $\gamma \gamma$ opacity, 
we adopt the standard model of Franceschini et al. (2008).

\section{Physical quantities in 3C~84}

In this section, we discuss physical parameters of 3C~84.
All of the quantities are summarized in Table~1.

\subsection{Total power of the jet : $L_{j}$}

The mass of the black hole in NGC~1275 is estimated to
be around $M_{\odot}\approx 8 \times 10^{8}M_{\odot}$ 
by gas kinematics  (Schawachter et al. 2013).
Correspondingly, the Eddington luminosity of NGC 1275 is
$L_{\rm Edd}\approx 1 \times 10^{47}~{\rm  erg~s^{-1}}$.
It is natural to suppose that 
the total power of the jet ($L_{j}$) 
satisfies  the relation of $L_{\rm Edd}\ge L_{j}$
\footnote{see however Ghiselleni et al. (2014)}.

Various estimates of the total power of the jet in 3C~84
at the central parsec region may be found in the literature.
 Heinz et al. (1998) argued that
 the time-averaged total power of the jet in NGC 1275
 probably exceeds 
  $L_{j}\sim 10^{46}~{\rm erg~s^{-1}}$.
  They derived the conclusion based on the 
  observed properties of X-ray cavities in the central region
  of the Perseus cluster, which is supposed to be inflated by
  relativistic particles of the shocked jet.
  They also suggested that 
  the jet power in a 
  quiescent state (corresponding to off-state mentioned
  in the paper of Reynolds and Begelman 1997)
  may be lower than $\sim 10^{46}~{\rm erg~s^{-1}}$.
Such a quiescent state case  would 
 correspond to the estimate of the power 
 $L_{j}\sim 5\times10^{44}~{\rm erg~s^{-1}}$
 in Abdo et al. (2009).
The observed luminosity at each energy band is 
of order of $\sim 10^{43}~{\rm erg~s^{-1}}$
from radio to GeV $\gamma$-ray band (Abdo et al. 2009).
Therefore, the bolometric luminosity is estimated 
to be $\sim 10^{44}~{\rm erg~s^{-1}}$.
Hence, a typical case of  
radiative efficiency of non thermal electrons 
with a few percent
results in the electron kinetic power 
of the order of  ${\rm a~few}\times10^{45}~{\rm erg~s^{-1}}$.
The proton component is also supposed to contribute to
the total jet power.
It is quite reasonable that 
$L_{\rm j}$ corresponds to a few percent of $L_{\rm Edd}$.
These estimates are consistent with a jet luminosity of the order of
 ${\rm a~few}\times 10^{45}~{\rm erg~s^{-1}}$,
 in agreement with the range of values reported in the literature.
Therefore, following Heinz et al. (1998),
we adopt 
$  L_{j} = 5 \times10^{45}~{\rm erg~s^{-1}} $
(i.e., 5\% of the Eddington power)
as a fiducial value.
%

\subsection{Fading radio lobe}

In the present work, we will focus on 
VLBA data at 15~GHz because there are sufficient archival data 
(MOJAVE project summarized by Lister et al. 2009, see also
http://www.physics.purdue.edu/MOJAVE/)
that enable us to explore basic properties of 3C~84 in detail
and the spatial resolution is suitable for exploring shells.

\subsubsection{ VLBA images of  fading radio lobe at 15~GHz}

In order to see basic characteristics of the radio lobe, 
we analyzed three epochs of relatively good quality 
VLBA archival data of 3C~84 at 15~GHz obtained in 
1994, 
2010, and 
2015 
(the project ID are
BR003, BL149CX, and BL193AS, respectively).
Two of them are adopted from MOJAVE data
(http://www.physics.purdue.edu/MOJAVE/).
The initial data calibration was performed with the Astronomical
Image Processing System (AIPS) developed at the
National Radio Astronomy Observatory. 
First, a priori amplitude
calibration was applied using the measured system noise
temperature and the elevation-gain curve of each antenna. 
We then calibrated the amplitude part of bandpass characteristics
at each station using the auto-correlation data. 
We applied opacity correction due to the atmospheric attenuation, 
assuming that the time variation of the opacity is not significant 
during each observation. 
The visibility phase and delay offset between different sub-bands
were solved by using 3C~84 itself.
Fringe fitting was performed with the AIPS task 
{\it "fring"} 
on 3C~84 by averaging over all the IFs.
Imaging and self-calibration were performed
using the Difmap software package (Shepherd 1997). 
The final images were produced after iterations of CLEAN, 
phase, and amplitude self-calibration processes.
We used a natural weighting scheme.

\begin{figure} 
\includegraphics
[width=16cm]
{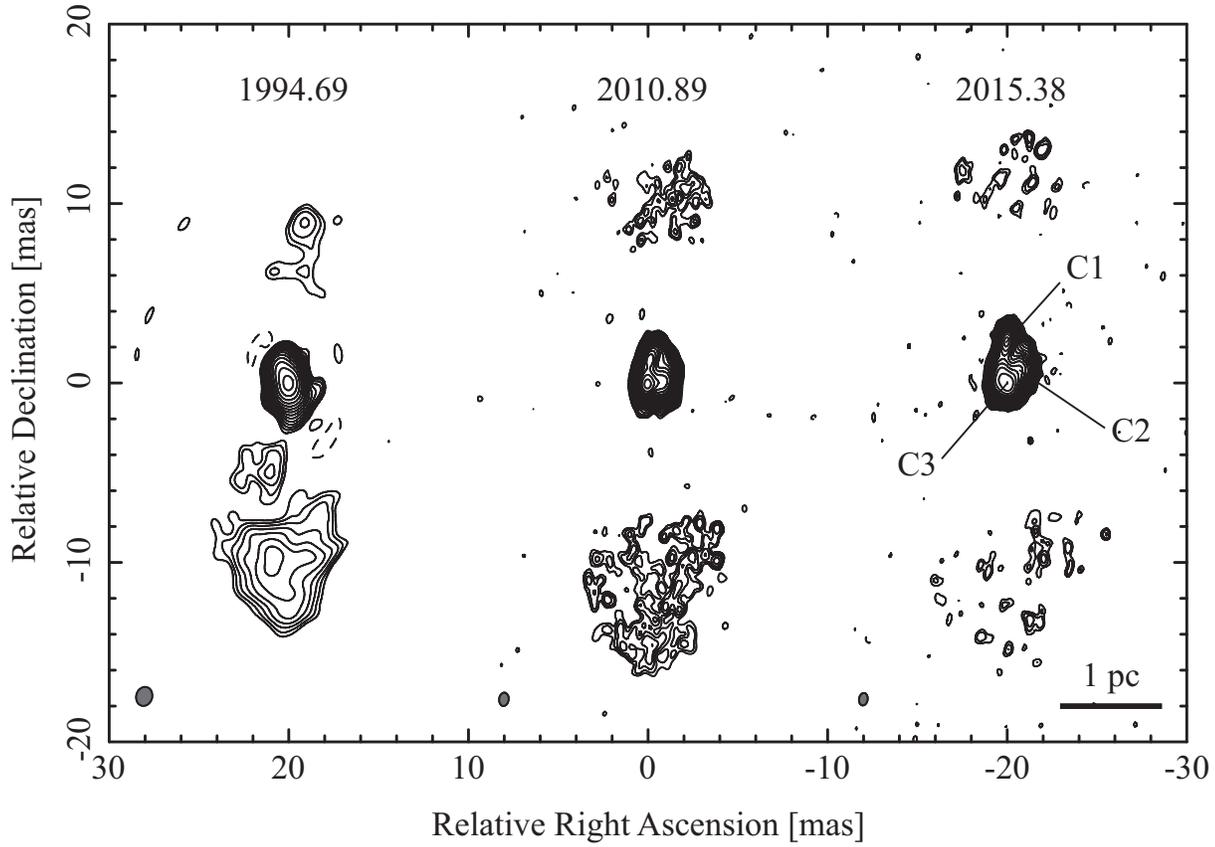}
\caption
{The comparison of the three epochs overall 3C~84 radio lobe images  
with VLBA at 15~GHz in 
1994, 
2010, and 
2015 
(data are adopted from the VLBA archival data with the project ID
BR003, BL149CX, and BL193AS, respectively).
The central compact lobe and a pair of fading radio lobes are seen 
in each epoch.
The total fluxes of the northern lobe, 
the central compact region (composed of C1, C2 and C3 components), 
and the southern lobe
are respectively,
0.75, 13.89, 6.66 Jy (in 1994),
0.52, 20.56, 1.44 Jy (in 2010), and 
0.40, 28.06, 0.76 Jy (in 2015).
Note that the brightness peak (phase-center) in the image
coincides with C1 in 1994 while the peak is at C3 in 2010 and 2015. 
The image rms is $\sim 3$~mJy/beam  (1~$\sigma$)
for the epoch in 2015.
}
\label{fig:15G}
\end{figure}

 In Figure~\ref{fig:15G}, we show 
 the obtained intensity map of the fading radio lobe.
The image root-mean-square (rms) of VLBA at 15~GHz
in Figure~\ref{fig:15G} 
of each three epochs  ($1~\sigma$) is, 
 18.5~mJy/beam (1994, ID BR003),
 3.1~mJy/beam (2010, ID BL149CX), and 
 3.0~mJy/beam (2015, ID BL193AS), respectively.
The high rms and large beam size for 
the observation performed in 1994( ID BR003)
is due to problems at the SC station which cause the 
flagging of the data from that antenna.
The  total flux of the fading lobe in each epoch 
is shown in the figure caption.
We define the source radius $R$ as 
the de-projected distance between the nucleus (C1 component)
to the head of the radio lobe and it is a measured quantity.
The projected angular distance is about 20~mas.
The allowed de-projection distance is  
$R_{\parallel} \approx  9-16~  {\rm pc}$
based on the previous estimate of the jet viewing angle 
$\theta_{\rm view}=25^{\circ}$ and $\theta_{\rm view}=49^{\circ}$
derived by Tavecchio et al. (2014) and  Fujita and Nagai (2017),
respectively.
At the center of each epoch, 
the bright central compact region consists of three sub-components: 
C1, likely hosting the source core, C2, a diffuse and faint component, 
and C3, a bright hot-spot-like component (see e.g., Nagai et al. 2010).
The central region accounts for the majority of the
source emission and its
synchrotron photons play a dominant role  as  seed photons
for IC scattering in the shell.

The long-term radio light curve of 3C~84
at 8~GHz shows the flux density increase started around 1960
(Nesterov et al. 1995).
Therefore, the age of the dying radio lobe in 3C~84
is estimated as $t\sim 50-60~{\rm years}$ in 2015.
In this work, we set $t=55$~{\rm years}.

\subsubsection{on the duration of the jet injection}

It is hard to estimate the duration of the jet injection ($t_{j}$) 
or equivalently the time when the jet stopped accurately,  
because it is not a direct observable.
At least, the fading of the radio lobe clearly indicates that
$t_{j}$ is shorter than $t$.
In this work, we will examine the three cases of
$t-t_{j}=$5, 10, and 30 ~{\rm years}.

\subsection{Shell}

Once we know the evolution of fading radio lobe,
then we can derive the advancing velocities of the shell.

\subsubsection{Advancing velocity parallel to the jet: $v_{\parallel}$}

The advancing velocity of the radio lobe head ($v_{\parallel}$) has been 
well constrained in the literatures and we simply follow it.
In  Asada  et al. (2006),
the authors measure it with the two epoch data 
(1998 August and  2001 August)
of VSOP observations at 5~GHz and
derived an advancing velocity as 
$v_{\parallel}\sim 0.5~c$.
Lister et al. (2013)
derived $v_{\parallel}\sim 0.3c$
in the framework of MOJAVE project.
The lobe advancing velocity 
seen in Figure~\ref{fig:15G} 
 is consistent with these $v_{\parallel}$
in the literature.


\subsubsection{Advancing velocity parallel to the jet: $v_{\perp}$}

On the contrary to $v_{\parallel}$,
little is known about 
propagation velocity of the shell perpendicular to the jet axis ($v_{\perp}$),
since there is no observational constraint on the high-$n$ shell.
In our model, same amount of internal energy is allocated 
the shocked torus and shocked ambient matter regions.
Therefore, the pressure in the shocked torus is larger
than that in the shocked ambient matter region because of 
the smaller volume.

The propagation speed of $v_{\perp}$
is governed by the balance between 
the ram pressure and the cocoon pressure in the perpendicular direction, 
i.e., $\rho_{\rm ext} v_{\perp}^{2} \propto P_{\rm c,\perp}$
where $P_{\rm c,\perp}$
is the pressure in the region behind the high-$n$ shell
(see Figure~\ref{fig:cartoon}).
The larger $\rho_{\rm ext}$
leads to
(1) the smaller $v_{\perp}$ and correspondingly
(2) the smaller radius for the region behind the high-$n$ shell ($R_{\perp}$).
Then, the pressure $P_{\rm c,\perp}$ plays an important role for determining $v_{\perp}$
because $P_{\rm c, \perp}\propto E_{\rm c, \perp}/V_{c, \perp}\propto R_{\perp}^{-2}$
where $V_{c, \perp}\propto R_{\perp}^{3}$ and  $E_{\rm c, \perp}\propto R_{\perp}$.
The smaller  $R_{\perp}$ leads to the larger $P_{\rm c, \perp}$.
Therefore, the larger  $\rho_{\rm ext}$ in the high-$n$ shell 
and the larger $P_{\rm c, \perp}$ are in the high-$n$ shell
largely cancelled each other out. 
Then, $v_{\perp}$ does not slow down significantly.
With the model parameters in the present study summarized in Table~1, 
we can obtain 
$v_{\perp}/v_{\parallel}
\approx R_{\perp}/R_{\parallel}
\approx 1/5$ for $t=55~{\rm years}$.

\subsubsection{ Magnetic fields strength: $B_{\rm shell}$}

Here we argue the viable range of magnetic field strength.
First, we argue the lower limit of magnetic field
strength averaged over the spatial scale $\sim 100$~pc.
Silver et al. (1998)  discovered  the extended radio halo structure at 330~MHz
(hereafter we call "milli-halo" according to their original naming)
with its averaged diameter of $\sim 500$~mas
which is slightly elongated along the orientation of
the jet axis.
Furthermore, the milli-halo has a brightness temperature 
$\sim 3\times 10^{7}$~K, it is surely non thermal.
Thus the emission most probably originated in the jet 
remnant generated by the past activity of the central engine.
Then Silver et al. (1998) estimate the magnetic field strength
in the milli-halo as 
$B_{\rm ext}\approx 200~{\rm \mu G}$
by assuming equipartition condition
together with the path length of 75~pc.
Taylor et al. (2006) estimated the field strength 
in  the central region ($<2$~kpc) of the Perseus cluster
by using the equipartition assumption between the
magnetic field and hot plasma which emit X-ray 
with a temperature of $5\times 10^{7}$~K and 
number density $\sim 0.3~{\rm cm^{-3}}$
and have obtained
 $B_{\rm ext}\approx 300~{\rm \mu G}$ which is comparable
 to the estimate obtained by Silver et al. (1998).
Taylor et al. (2006) further estimated the field strength 
within the $10$~pc scale radio lobe as 
by measuring the rotation measure (RM) associated 
with the lobe's tip, which can be regarded as the 
hotspot. The measured value shows 
$RM \sim 7\times 10^{3}~{\rm rad~m^{-2}}$.
By adopting the path length as 1~pc, they
derived the value of  $B_{\rm ext}\approx 50~{\rm \mu G}$.
Based on these previous works, here we  set
the  fiducial  magnetic field strength in the 
shell as $ B_{\rm shell} \approx 0.1~{\rm mG}$.

\subsection{External medium}

Hereafter, we denote the  number density of external medium as $n_{\rm ext}$.
We assume that the external medium consists of both a diffuse gas with low
number density  ($n_{\rm amb}$) 
 surrounding the radio source, and a circum-nuclear structure, e.g. a torus, of
dense ionized gas ($n_{\rm torus}$).
A sketch is shown in Figure~\ref{fig:cartoon}.

\subsubsection{Number density of diffuse ambient medium: $n_{\rm amb}$}

Taylor et al. (2006) estimated the number density
by using the deep Chandra observation (Fabian et al. 2006).
Within the central 0.8kpc,
the density profile is severely affected by
the nucleus. So, they estimated an average
central density over the inner 2kpc to be
 $n_{\rm amb}\sim 0.3~{\rm cm^{-3}}$.
Regarding the upper limit of number density 
we adopt the value recently obtained by Fujita et al. (2016).
They make two  assumptions that 
(1) hot gas outside the Bondi radius is in nearly a hydrostatic
equilibrium in a gravitational potential, and 
(2) the gas temperature near the galaxy
centre is close to the virial temperature of the galaxy.
Then, they obtain
$n_{\rm amb}\sim 10~{\rm cm^{-3}}$
in the inner part of Perseus cluster.
Then we obtain
$0.3~{\rm cm^{-3}}\le n_{\rm amb} \le  10~{\rm cm^{-3}}$
by regarding the value derived by Fujita et al. (2016) 
as the upper limit of $n_{\rm amb}$.
In this work, we set 
$n_{\rm amb}\approx 0.3~{\rm cm^{-3}}$
as a fiducial one based on Fabian et al. (2006).

\subsubsection{Number density of plasma torus: $n_{\rm torus}$}

The existence of  dense thermal gas surrounding
3C~84 was discovered by O'dea et al. (1984).
They suggested that 3C~84 is embedded in a dense thermal gas
with the  number density $\sim 2\times 10^{3}~{\rm cm^{-3}}$
derived from the 
condition of depolarization by differential Faraday rotation
in the external plasma.
Walker et al. (2000) also
confirm that the spectra are consistent with FFA
by VLBA observations  at 5, 8, 15, and 22 GHz
and the northern radio lobe
feature is on the far side of the system relative to the Earth.
Therefore, the size of the plasma (ionized gas)
 should be spatially extended to 
the scale at least comparable to the  radio lobe seen in Figure~\ref{fig:15G}
since the northern lobe is still hidden by the FFA.

Although the geometrical details of the thermal gas
are highly uncertain, 
it is clear that 
the required total power of the jet would be very large
if the dense thermal gas with $\sim 10^{3}~{\rm cm^{-3}}$
completely envelopes the overall 3C~84 system.
For this reason,
we  regard it as the dense ambient matter 
and as a geometrically thick plasma torus.
Here, we set 
$n_{\rm torus} = 1\times 10^{3}~{\rm cm^{-3}}$
with the path length of about $10~{\rm pc}$
which is comparable to $R$
(see the next sub-sections.)

\subsection{External photon field}

\subsubsection{Synchrotron emission from the central compact region}

As already explained, it is well known that the central compact region consists of three sub-components: 
C1, likely hosting the source core, C2, a diffuse and faint component, 
and C3, a bright hot-spot-like component.
Hereafter, synchrotron photons
coming from the whole central component region are considered, 
with no distinction among those sub-components. 
This should make the method and results of this work easier and clearer
to read and follow.

\begin{figure} 
\includegraphics
{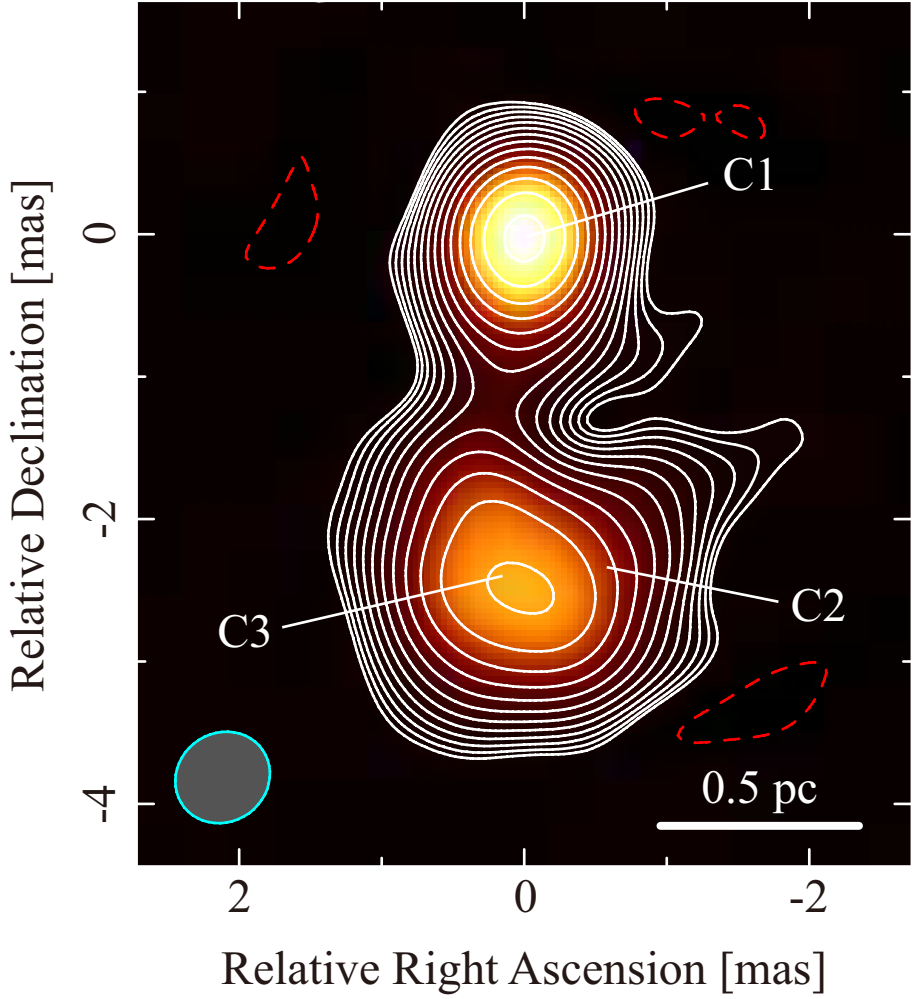}
\caption
{KaVA image of  the central  compact region in 3C~84 obtained in 2015.
The map peak intensity is 5.3 Jy/beam. 
The synchrotron emission from the central compact region 
is the major seed photons for the IC scattering in the fossil shells.
}
\label{fig:KaVA43GHz}
\end{figure}

In Figure~\ref{fig:KaVA43GHz}, 
we show the recent image of the central compact region with KaVA at 43~GHz .
KaVA is a combined VLBI array with  KVN (Korean VLBI Network) and 
VERA (VLBI Exploration of Radio Astrometry) operated by KASI
and NAOJ, respectively
and its baseline lengths range from 305 to 2270 km
(see Niinuma et al. 2014 for details).
This is one of the epochs of a long-term KaVA monitoring of 3C~84 at 43GHz
(http://radio.kasi.re.kr/kava/) and 
we conducted a data reduction in a standard way, 
already described in sub-section 3.2.1.
The obtained total flux of the central compact region  at 43~GHz is about 15~Jy in 2015
and thus we obtain 
$L_{43G}=4\pi D_{\rm L}^{2} S_{\nu}\nu \approx 4 \times 10^{42}~{\rm erg~s^{-1}}$.
The measured total flux density  of the
central compact region is consistent with the 
one obtained by the blazar monitoring project led
by Boston university group
(https://www.bu.edu/blazars/VLBAproject.html).

The long termVLBI monitoring of the central compact region at 22~GHz shows
a gradual monotonic increase of the flux from $\sim$2006
 (Nagai et al. 2010, 2012; Suzuki et al. 2012; Chida et al. 2015).
The total flux of the central compact region at 14mm measured by VERA is 
$\sim$20~Jy and from this we obtain
$L_{22G}=4\pi D_{\rm L}^{2} S_{\nu}\nu \approx 3 \times 10^{42}~{\rm erg~s^{-1}}$.
In the present work, we set an averaged spectral index as
$\alpha_{R}\approx 0.3$ in 
$S_{\nu}$ of the central compact region (Chida et al. 2015).
In \S 4, we will find that 
the synchrotron photons from the central compact region 
are the dominant seed photons for the IC scattering.

\subsubsection{Synchrotron emission from the fading radio lobe}

Synchrotron photons from the fading radio lobe 
are also regarded as seed photons for IC in the shell
although it does not play a dominant role.
As shown in Figure~\ref{fig:15G}, the total fluxes of the 
fading radio lobes at 15~GHz
are 7.4, 2.0, 1.2~Jy in 1994, 2010 and 2015,
respectively.
Then the corresponding luminosity of the fading radio lobe
in 2015 is 
$L_{\rm lobe} \approx 1\times 10^{41}~{\rm erg~s^{-1}}$
and it is shown in Table~1.
In \S 4, 
the synchrotron photons from the fading radio lobe 
will turn out to be 
less dominant as seed photons for the IC scattering.

\subsubsection{Thermal emission from accretion disk and  dust torus}

Here we estimate accretion disc luminosity ($L_{UV}$) in 3C~84
using the observed line spectra.
Using the observation data obtained by Kanata telescope’s HOWPol 
(Hiroshima One-shot Wide-field Polarimeter) (Kawabata et al. 2008) ,
Yamazaki et al. (2013) addressed 
the variation of the line ratio 
of (H $\alpha$+[NII])/[OII]  which reflects 
the activity of UV photons from accretion disc.
The  broad H $\alpha$ line shows
no significant variability during 2010-2011.
Although there is a possibility of a long-term ($\sim$10 year scale)
change of the activity of the accretion disk,
here we estimate  $L_{UV}$  by regarding
the flux obtained by Kanata as a conservative minimum value.
Using the same data, we further revisited and
evaluated the line spectra in 3C~84 and
we estimate the UV luminosity as follows (Kino et al. 2016). 
The [OII] luminosity is derived as $L_{\rm [OII]}=1.9\times 10^{42}~{\rm erg~s^{-1}}$ by Ho et al. (1997),
while the H alpha plus [NII] line luminosity 
measured by KANATA HowPOL is
$L_{H\alpha+[NII]}=1.2\times 10^{42}~{\rm erg~s^{-1}}$.
According to Ho et al. (1997),
$L_{[NII]}$ typically contributes to
$L_{H\alpha+[NII]}$ up to  $\sim 20\%$.
Adopting the  value of [OI] luminosity from Ho et al. (1997)
and the empirical relation by Greene and Ho (2005),
we obtain $L_{\rm UV}\approx 5 \times 10^{42}~{\rm erg~s^{-1}}$.
\footnote{
Ho et al. (1997) did not explicitly mention the value of $H_{0}$ in their paper.
If they adopted  $H_{0} = 50~{\rm km/s/Mpc}$, 
then they underestimated of line luminosities  by a factor of $\sim 2$.
}
In addition, the luminosity ratio of
$L_{\rm bol}/L_{\rm Edd}\sim 4\times 10^{-3}$
where $L_{\rm bol}$ is the bolometric luminosity of 
3C~84 (e.g., Levinson et al. 1995 and reference therein).
Thus, the classification of the accretion flow in 3C~84 may  
settle down on the border between the standard 
Shakura-Sunyaev disk (Shakura and Sunyaev 1973)
and Radiatively Inefficient Accretion Flow (RIAF).
In the present work, however, 
we do not take the RIAF emission into account
because  it is negligibly small as seed photons for IC
at the fossil shell.

The dust emission at the center of NGC~1275
has been investigated with {\it Herschel}  data 
by Mittal et al. (2012) and they derived 
the total dust luminosity 
as $L_{\rm IR, total}\sim 5\times 10^{44}~{\rm erg~s^{-1}}$.
spatially- ntegrated over arcsec angular-size scale.
Because of heavy absorption by  dust torus in young compact radio sources
(e.g., Kawakatu et al. 2009b;  Ostorero et al. 2010),
it is difficult to estimate the dust-torus luminosity accurately.
Hence, we use the work of Calderone et al. (2012)
to give us a better estimation of the dust torus luminosity ($L_{IR}$).
Calderone et al. (2012) explore the fraction
of torus re-emission of absorbed accretion disc radiation for
about 4000 radio-quiet AGNs and they found that the dust torus
reprocesses  1/3-1/2 of the accretion disk luminosity. 
Based on their work, we set $L_{\rm IR, torus} =\frac{L_{UV}}{2}$.
The lower value of $L_{\rm IR, torus}$ 
compared to  $L_{\rm IR, total}$ obtained by Mittal et al. (2012) 
can be consistent with each other, since the spatial resolution of {\it Herschel}
is much larger than the size of dust-torus (sub-arcsec scale)
and it probably contains galactic-dust emission.

\section{Predicted shell spectra}

In the previous section,
we carefully discuss 
the observed quantities of 3C~84 and
the physical quantities in the model, 
which are summarized in Tables 1 and 2.
Since 3C~84 is one of the best-studied radio sources,
we have  tight constraints  of observational quantities.
In this section, we show the non-thermal emission spectra
from the fossil shell in 3C~84 after stopping the jet energy injection.
Here, we conservatively
set the total luminosity of the central compact region
as constant in time. 
Since the central compact region still gets brighter, the resultant fossil shell
spectra to be shown here would correspond to conservative lower limit cases.

As for treatment of fading radio lobes,
we simply follow our previous work of I15.
We set a large value of electron gyro-factor
in the lobe as $\xi_{e, \rm lobe}=10^{7}$.
In general, the gyro-factor is proportional to a particle
acceleration timescale  and thus
it determines the maximum energy of those accelerated particles.
Change of $\xi_{e, \rm lobe}$ value has no impact on the results of this work.
We set the power-law index of injected electrons in the lobe
as $p_{e,\rm lobe}=2.2$ corresponding to the standard
value for relativistic shocks 
(e.g., Bednarz \& Ostrowski 1998; Achterberg et al. 2001; Kirk 2000).
As for energetics, we simply assume the 
equipartition between $B$-fields and electrons 
i.e., $\epsilon_{e}=\epsilon_{B}\approx 0.1-0.01$. 
For avoiding complexity of the figures,
we do not overlay the fading radio lobe spectra.
We find that the IC emission due to
the synchrotron seed photons from the fading lobe are
not dominant in the fossil shell spectra.

We also note that
an actual total emission from the fossil shell 
should be the sum of the high-$n$ and low-$n$ spectra.
Below, we separately  discuss the spectra from high-$n$ and low-$n$
 for better clearness.

\subsection{The case of  $n_{\rm ext}=n_{\rm amb}$}

\begin{figure} 
\includegraphics
[width=15cm]
{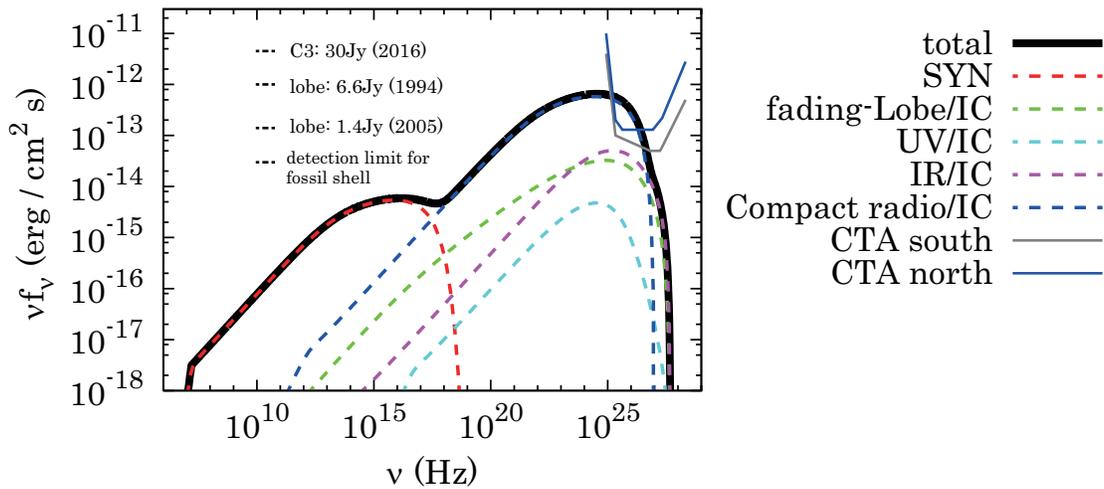}
\caption
{Low-$n$ shell spectrum in 3C~84 (predicted in 2015 and evolution after that)
with 
$n_{\rm amb}=0.3 ~{\rm cm^{-3}}$,
$L_{\rm j}=5\times 10^{45}~{\rm erg~s^{-1}}$,
$t=55~{\rm yr}$
and $B_{\rm shell}=0.1~{\rm mG}$.
Further model parameter values and the observed quantities are completely
summarized in Tables \ref{table:lobe} and \ref{table:environment}.
The shell spectrum is IC-cooling dominated and 
the IC component of synchrotron photons
from the central compact region (gray line) is dominant.
}
\label{fig:low-n}
\end{figure}

In Figure~\ref{fig:low-n}, we show the emission spectrum
from the shocked ambient medium (i.e., low-$n$ shell).
Following the physical parameters reported in 
Tables \ref{table:lobe} and \ref{table:environment}
and discussed in \S 3,
we estimate the expected emission from the 
shell that is expanding in a low-density
ambient medium, i.e., $n_{\rm amb}=0.3 ~{\rm cm^{-3}}$.
The jet power, age, and the magnetic field strength in the shell are, respectively,
 $L_{\rm j}=5\times 10^{45}~{\rm erg~s^{-1}}$,
 $t=55~{\rm years}$ and 
 $B_{\rm shell}=0.1~{\rm mG}$.
By solving the evolution governed by Eq.~(\ref{eq:R})
with these $L_{\rm j}$, $t$  and $n_{\rm amb}$,
we obtain $R_{\parallel} \approx 8~{\rm pc}$ 
which is similar to the minimum value of the estimated
$R_{\parallel}\approx 9~{\rm pc}$.
(Since a spherical symmetry is assumed Eq.~(\ref{eq:R}) for simplicity, 
all of the jet power is isotropically ejected.
Hence, a slightly smaller $R_{\parallel}$ by the model is 
more consistent.)
The corresponding advancing velocity of the shell
is $v_{\perp}\sim 0.26~c$.
In addition, we can readily find that 
a larger $R_{\parallel}$ requires 
a fairly large $L_{\rm j}$.

The shell spectrum is IC-cooling dominated and 
the IC peak occurs in  the TeV $ \gamma$-ray energy band. 
The IC component  of synchrotron photons
from the central compact region in Figure~\ref{fig:15G})
is dominant in the  shells.
The distance from the seed photon source 
and the fossil shell is taken into account. 
Even though the distance from the central compact region
and the fossil shell is about 10~pc and the photon energy
density decrease as $R^{-2}$, the IC component 
of the central compact region is still more dominant
than the IC component of the fading radio lobe
simply because the synchrotron luminosity of the 
central compact region is much higher than that from the fading lobe.

The predicted shell spectra have a trend similar to
the ones shown in our previous work of  I15.
The opacity
for  $\gamma \gamma$ interaction between EBL and TeV photons
is sufficiently small  because of its proximity
and we find that the predicted TeV $\gamma$-ray flux 
can be comparable to the sensitivity of CTA
(https://web.cta-observatory.org/science/cta-performance/)
with the integration time 50~hours.
As reported in the Astronomer's telegram
(http://www.astronomerstelegram.org), 
the data show variabilities in TeV $\gamma$-ray flux
(e.g.,  
Mirzoyan 2016, 2017; 
Mukherjee and VERITAS Collaboration 2016, 2017;
Lucarelli et al. 2017; 
Ahnen et al. 2016).
Such  variabilities can be naturally explained by
the emission from the blazar region (e.g., Tavecchio et al. 2014).
On the other hand, 
the fossil-shell emission does not show such variabilities.
Since the TeV $\gamma$-ray  emission from a fossil-shell is 
less luminous than those from the blazar region,
a low-activity phase of the blazar region is favored 
in search of the fossil shell emission.

It is well known that
IC scattering process is divided into two regimes, i.e.,
Thomson and Compton regimes.
Using the characteristic energy of 
the seed photons ($E_{\rm seed}$)
and electrons
($E_{e}=\gamma_{e}m_{e}c^{2}$),
the following relation is satisfied
\begin{eqnarray}
2 E_{\rm seed}\times E_{e}\left\{ \begin{array}{ll}
\ll     (m_{e}c^{2})^{2} & {\rm (Thomson~regime)}\\
\gtrsim (m_{e}c^{2})^{2} & {\rm (Compton~regime).}\\
\end{array} \right. 
\end{eqnarray}
The maximum energy of IC spectrum ($h \nu_{\rm IC, max}$)
in the case of the Thomson regime is given by
\begin{eqnarray}
h \nu_{\rm IC, max} \approx \gamma_{e,\rm max}^{2} E_{\rm seed} ~{\rm (Thomson~regime)} ,
\end{eqnarray}
while $h \nu_{\rm IC}$ is written as
\begin{eqnarray}
h \nu_{\rm IC, max} &\approx& 
\gamma_{e,\rm max} m_{e}c^{2}   \nonumber \\ 
&=& E_{e,\rm max}   ~{\rm (Compton~regime),}
\end{eqnarray}
for the Compton  regime.
Keeping this in mind,
let us argue which scattering regime is realized for seed photons
considered in this work.
Against UV photons with the energy $E_{\rm UV}$  from an accretion disk,
the IC scattering is taken place in Compton regime 
for electrons with their energy
\begin{eqnarray}
E_{e}\gtrsim
\frac{(m_{e}c^{2})^{2}}{2E_{\rm UV}} 
\approx 12~{\rm GeV}
\left(\frac{E_{\rm UV}}{10~{\rm eV}}\right)^{-1}  .
\end{eqnarray}
Thus we find that
$h \nu_{\rm IC, max}$ is significantly limited by the Klein-Nishina effect
although $E_{e,\rm max}/m_{e}c^{2}$
extends up to $\sim10^{7.5}$  in this case.
From this, we can easily understand that IC scattering against 
IR photons from dusty torus is also in the range of Compton regime.
On the other hand, 
the seed photons from the central compact region
with its characteristic energy $E_{\rm mm}$
is IC scattered in the Thomson regime and thus we obtain
\begin{eqnarray}
h \nu_{\rm IC, max} \approx 
10^{12}~{\rm eV}
\left(\frac{\gamma_{e,\rm max}}{10^{7.5}}\right)^{2} 
\left(\frac{E_{\rm mm}}{10^{-3}~{\rm eV}}\right)   .
\end{eqnarray}
This well explains the IC spectra in 
Figure~\ref{fig:low-n}.

\begin{figure} 
\includegraphics
[width=15cm]
{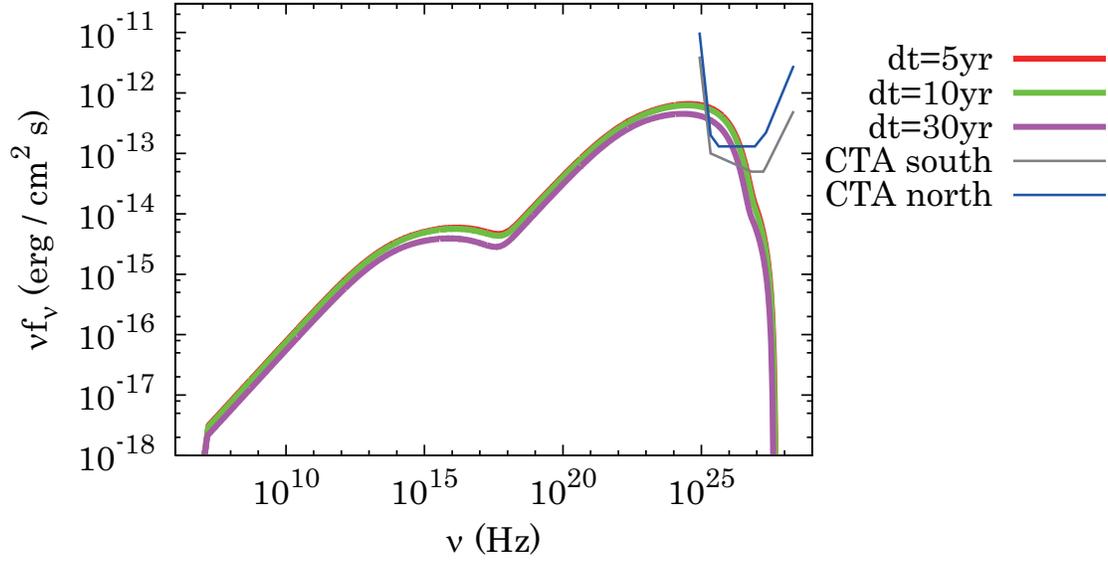}
\caption
{Spectral evolution of the fossil shell 
where $n_{\rm ext}=n_{\rm amb}=0.3~{\rm cm^{-3}}$.
Three epochs with the durations after the jet stopping
as 5, 10 and 30 years are shown here.
The spectra are almost constant in time 
because of continuous injection of fresh electrons in the shell
via the forward shock.}
\label{fig:evolution-low-n}
\end{figure}

In  Figure~\ref{fig:evolution-low-n}, we show the three epochs of 
the predicted shell spectra.
As already explained in I15,
the shell almost keeps its emission flux level 
because fresh electrons are continuously
supplied into the shell via the forward shock
driven by the cocoon/radio-lobes 
while the radio-lobes rapidly fade out
without jet energy injection.

\subsection{The case of $n_{\rm ext}=n_{\rm torus}$}

\subsubsection{The case  with B-amplification}

Here, we take the process
non-linear amplification of the magnetic fields
into account (Lucek \& Bell 2000; Bell \& Lucek 2001),
although little attention has been paid in the research 
field of AGN jets so far.
Based on Lucek \& Bell, it is expected that cosmic ray (CR) streaming
 in AGN jets would drive
large-amplitude Alfvenic waves and the CR streaming energy is
transferred to the perturbed magnetic field of the Alfven waves. 
Let us discuss 　the case when 
B-amplification is effective in the torus region
(i.e., $n_{\rm ext}=n_{\rm torus}$). 
In this work, following the pioneering work of 
Berezhko (2008) which first introduces nonlinear 
amplification process of the magnetic field 
in AGN jets into account, we employ the empirical relation of 
\begin{eqnarray}
\frac{B_{\rm shell}^{2}}{8\pi} 
\approx 3 \times 10^{-3} \rho_{\rm torus}v_{\perp}^{2}, \nonumber
\end{eqnarray}
obtained by Berezhko (2008)
where  $\rho_{\rm torus}=n_{\rm torus}m_{p}$.
Then, we conservatively obtain the maximum value of the  high-n shell's field as
\begin{eqnarray}
 B_{\rm shell} \lesssim 17~{\rm mG},
\end{eqnarray}
where $n_{\rm torus}=1\times 10^{3}~{\rm cm^{-3}}$ and
$v_{\perp}\sim v_{\parallel}/5\approx 0.05c$
are adopted here.
This empirical relation by Berezhko (2008) is justified by recent studies of 
high-resolution Magneto-hydrodynamical (MHD) simulations. 
In particular, turbulence 
can significantly amplify $B$-field 
(e.g., 
Giacalone \& Jokippi 2007;
Inoue et al. 2009; 
Guo et al. 2012;
Sano et al. 2012;
Fraschetti 2013;
Ji et al. 2016).
For instance, the amplification factor of $B$-fields
obtained by Ji et al. (2016) indeed reaches  over $\sim 100$ 
which is  consistent with the value discussed here. 
Note that a pile-up process of the $B$-field lines 
as the jet propagates sweeping the field lines 
in the torus could also help $B$-fields amplification
(Rocha da Silva et al. 2015).

\begin{figure} 
\includegraphics
[width=15cm]
{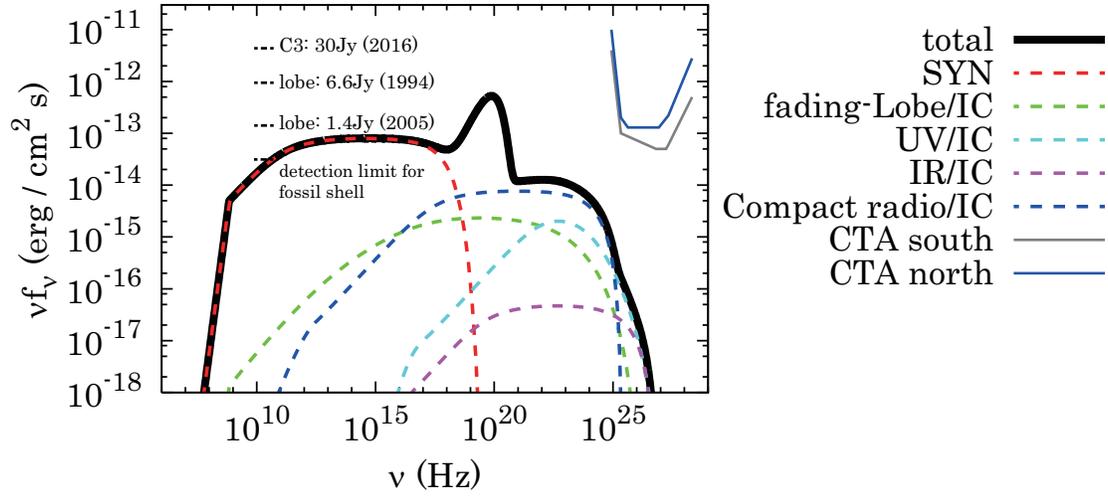}
\caption
{Same as Figure~\ref{fig:low-n} but with 
$n_{\rm ext}=n_{\rm torus}=1\times 10^{3}~{\rm cm^{-3}}$
and $B$-field amplification.
This case  is defined as high-$n$ shell.
The bump at $10^{20}$~Hz corresponds to thermal bremsstrahlung emission.}
\label{fig:high-n-Bamp}
\end{figure}

In Figure~\ref{fig:high-n-Bamp}, we present the resultant 
fossil-shell emission spectra of  dense fossil shell
when the amplified B-fields  become $ B_{\rm shell} \sim 17~{\rm mG}$.
The bump at $10^{20}$~Hz corresponds to thermal bremsstrahlung emission
from the fossil-shell.
The temperature and number density in the fossil-shell are
determined by Rankine-Hugoniot conditions (see I15 for details).
In radio band,
a bright synchrotron emission from the dense fossil shell
is predicted.
Here we estimate a typical detection-limit against 
a high-$n$ shell at 15~GHz for future VLBI observation.
As already shown in Figure~\ref{fig:15G}, 
the typical image rms is　found as 3mJy/beam.
Here, the case of  7-$\sigma$ detection is considered.
A required number of the VLBA beam at 15~GHz 
which can fill the the shell-surface area on the sky plane can be 
approximately estimated as
$\left[2\pi f_{\rm torus}\times 4~{\rm mas}(4~{\rm mas}\frac{\delta R}{R})\right]
  /\left[((\pi/4)(0.7\times 0.4)~{\rm mas}\right] \sim  19$
where $20/5=4{\rm mas}$ as the angular size of $R_{\perp}$, 
 $f_{\rm torus}=1/2$ and  $\delta R/R= 1/12$  are used. 
Then we can get the typical detection-limit  as 
$3~{\rm mJy/beam} \times 19~{\rm beam} \times 7\sigma
\approx 2 \times 10^{-14}~{\rm erg~s^{-1}~cm^{-2}}$.
By comparing  this   detection-limit and the predicted shell spectra,
we can argue a detectability of the shell emission  at 15GHz.
The predicted flux density is comparable  with 
the  detection-limit at 15~GHz. 
Therefore, performing 
deeper imaging observations of VLBA and/or other VLBIs 
in the future will generally give us meaningful 
constraints on fossil shell model parameters.
For example,
a usage of High Sensitivity Array (HSA), 
which is VLBA together with
the Green Bank Telescope, phased VLA, and Effelsberg 
(https://science.nrao.edu/facilities/vlba/proposing/HSA)  
would enable us to get better sensitivity by an order of magnitude
and will increase the chance to detect the fossil shell emission in 3C~84.

It is worth mentioning that the detectability of the shell 
might be limited not by  the image thermal noise but the dynamic range.
In such a case, better uv-coverage is essential for the detection.
The total flux  of the central compact region 
is about $\sim 10$~Jy level with a year-scale increase.
Then, the required dynamic range is about
a few times of 1000
if the flux density of the fossil shell is milli-Jy level.
Hence, the required dynamic range
can be attainable with typical/normal VLBA observations 
 which can reach about a few times of 1000 (e.g., Perley 1999).



\subsubsection{Without B-amplification}

\begin{figure} 
\includegraphics
[width=15cm]
{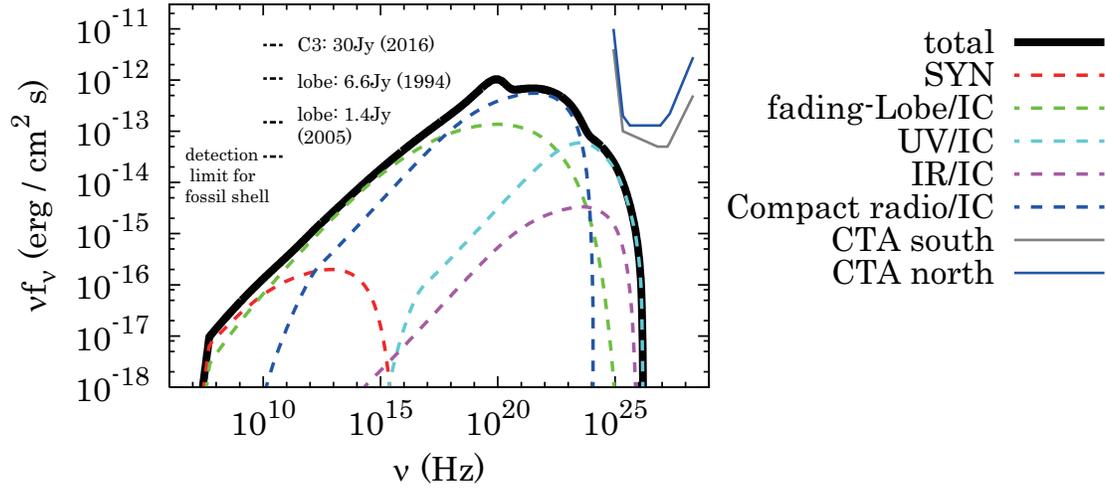}
\caption
{Same as Figure~\ref{fig:low-n} but with 
$n_{\rm ext}=n_{\rm torus}=1\times 10^{3}~{\rm cm^{-3}}$.,
i.e., without $B$-field amplification.
}
\label{fig:high-n}
\end{figure}

In Figure~\ref{fig:high-n}, we show the emission spectra
for the case of high-$n$ shell with  $n_{\rm ext}=n_{\rm torus}$.
The number density of the torus  is
$n_{\rm torus}=1\times 10^{3}~{\rm cm^{-3}}$ 
which is based on the constraint obtained by  O'Dea et al. (1984).
The other values of model parameters and the observed quantities are
the same as the ones in Figure~\ref{fig:low-n}, i.e., 
 $L_{\rm j}=5\times 10^{45}~{\rm erg~s^{-1}}$ and 
 $B_{\rm shell}=0.1~{\rm mG}$
and $\epsilon_{e}=0.1$.

Compared with the case of low-$n$ shell spectra, 
 IC component is less luminous in TeV $\gamma$-ray energy band.
 This is due to the frequency-peak shift 
of the two IC components whose seed photons are
the central compact region and 
the fading radio lobe.
This is caused by 
(1) the shift of 
$E_{e,\rm max}\sim 10^{5.5}m_{e}c^{2}$ 
due to the IC cooling, and 
(2) decrease of the velocity which leads to the longer 
timescale of electron acceleration since $t_{e,\rm acc}\propto v^{-2}$.
Then, the peak of IC against seed photons from the central compact region
in this case is in the Thomson regime which can be written as 
$h \nu_{\rm IC, max} \approx 
10^{9}~{\rm eV}
\left(\frac{\gamma_{e,\rm max}}{10^{5.5}}\right)^{2} 
\left(\frac{E_{\rm mm}}{1~{\rm meV}}\right)$.
As for IC scattering against UV and IR seed photons,
$E_{e,\rm max} \gg 12~{\rm GeV}$ still realizes
Compton regime at higher energy range.


\section{Implication for cosmic-ray proton production}

\subsection{UHECR production in the dense plasma torus?}

The origin of ultra high-energy cosmic ray (UHECR)
with energies above the ankle of
$\gtrsim 10^{18.5}$~eV is still under debate 
(e.g., 
Nagano \& Watson 2000;
Kotera \& Olinto 2011).
A jet in AGN is generally expected as one of 
the most plausible sites for UHECR production 
(e.g., Takahara 1990; 
Rachen \& Biermann 1993;
Norman et al. 1995;
Takami \& Horiuchi 2011;
Kino \& Asano 2011;
Murase et al. 2012).
It is suggested that UHECR production 
at the nucleus regions of AGNs tends to cause
problems due to various energy loss processes
(e.g., Pe'er et al. 2009; Pe'er and Loeb 2012). 
Therefore, it is worthwhile discussing a new possibility 
away from the central nucleus regions of AGNs.

Since higher value of $B$-fields can be expected 
if $B$-field amplification process is in action in dense environments,
here we discuss a feasibility of UHECR production in high-$n$ shells
for the first time.
For simplicity,
(1) we assume the standard diffusive 
shock acceleration 
(e.g., Blandford and Eichler 1987), and 
(2) we neglect the possible existence of  heavy nuclei.
It is well known that the maximum accessible energy 
of UHECRs ($E_{p,\rm max}$)
is governed by both 
the confinement condition and 
energy-loss/escape condition
(e.g., Kotera and Orinto 2011 and references therein).

The Larmor radius of  UHECRs ($r_{\rm L}\equiv E_{p}/e B_{\rm shell}$)
should be smaller than the shell width 
i.e., $r_{\rm L}\gtrsim  \delta R_{\perp}$.
The typical $r_{\rm L}$ for high-$n$ fossil shell is given by
\begin{eqnarray}
r_{\rm L}\approx 
0.13~{\rm pc}
  \left(\frac{E_{p}}{2 \times 10^{18}~{\rm eV}}\right)
  \left(\frac{B_{\rm shell}}{17~{\rm mG}}\right)^{-1}  ,
\end{eqnarray}
or equivalently,
\begin{eqnarray}\label{eq:E_max}
E_{p}  \leq E_{p,\rm max} \approx
2 \times 10^{18}~{\rm eV}
  \left(\frac{\delta R_{\perp}}{0.13~{\rm pc}}\right)
  \left(\frac{B_{\rm shell}}{17~{\rm mG}}\right)^{+1}  ,
\end{eqnarray}
where 
we use $\delta R_{\perp} \sim 1.6~{\rm pc}/12\approx 0.13~{\rm pc}$.
Essentially, the confinement condition governs
the maximum energy of the UHECR  ($E_{p,\rm max}$).

The acceleration timescale of protons ($t_{p,acc}$) generally satisfies
 $t_{p,acc}\lesssim  {\rm min}[t, t_{\rm diff}, t_{\rm loss}]$
where 
$t_{\rm diff}$and
$t_{\rm loss}$ are 
an diffusive escape timescale of UHECRs from the  acceleration region and
an energy-loss timescale of UHECRs in the  acceleration region, respectively
(e.g., Norman et al. 1995).

The acceleration
timescale of protons 
at the high-$n$ fossil shell in the coasting phase is given by
\begin{eqnarray}
 t_{p,acc} = 
 0.42~{\rm yr}
 \left(\frac{E_{p}}{2\times 10^{18}~{\rm eV}}\right)
 \left(\frac{B_{\rm shell}}{17~{\rm mG}}\right)^{-1}
  \xi _{p}  ,
\end{eqnarray}
where $\xi_{p}$ is the gyro-factor
for proton accelerations at the  shell.

Following the study of Gabici et al. (2009),
we adopt the diffusion coefficient (Eq. (13)) as
$ D_{p} =
 1\times 10^{28}
 \chi
 \left(\frac{E_{p}}{10^{10}~{\rm eV}}\right)^{1/2}
 \left(\frac{B}{3\mu G}\right)^{-1/2}
 ~{\rm cm^{2}~s^{-1}}$
where $\chi$ is a parameter which expresses
deviations from the average Galactic diffusion 
coefficient.
The  value of $\chi$  is highly uncertain.
The case of $\chi=1$ is identical to the 
case of the diffusion in Galactic interstellar medium.
The case $\chi<1$ accounts for a
possible suppression. The value of $\chi$ will
depend on the power spectrum of magnetic field turbulence
(e.g., Gabici et al. 2009 and reference therein).
Then, a typical diffusion timescale of the UHECRs can be estimated as
$ t_{\rm diff} =
 \frac{\delta R_{\perp}^{2}}{6D_{p}}    
 \approx 4.7\times 10^{-4}~ {\rm yr} 
  \left(\frac{\delta R_{\perp}}
{0.13~{\rm pc}}\right)^{2}
\chi^{-1}
 \left(\frac{E_{p}}{2\times 10^{18}~{\rm eV}}\right)^{-1/2}
 \left(\frac{B_{\rm shell}}{17~{\rm mG}}\right)^{1/2}$
 (Eq.~(3) in Gabici et al. 2009).
In order to satisfy the condition of 
 $t_{p,acc}\le t_{\rm diff}$,
the value of $\chi\sim 10^{-3}$  is required.
Although little is known about $\chi$ in AGN torus,
highly turbulent condition 
in AGN torus of magneto-ionized gas is expected
(e.g., Wada et al. 2002). 
Thus, a much slower diffusion 
compared with the ordinary  interstellar medium
could be realized because of turbulence in the torus.

It is readily found that $t_{\rm loss}$ is not competitive to $t_{p, acc}$.
The timescales of proton synchrotron and $pp$ collisions are given by
\begin{eqnarray}
t_{p,\rm syn}\approx 2.5\times 10^{5}~{\rm yr}
 \left(\frac{B_{\rm shell}}{17~{\rm mG}}\right)^{-2}
  \left(\frac{E_{p}}{2\times 10^{18}~{\rm eV}}\right)^{-1}  ,
\end{eqnarray}
and 
\begin{eqnarray}
t_{pp}\approx 5\times 10^{4}~{\rm yr}
 \left(\frac{n_{\rm shell}}{10^{3}~{\rm cm^{-3}}}\right)^{-1} ,
\end{eqnarray}
respectively.
These are significantly longer than $t_{p,acc}$.
Therefore, we conclude that
the high-$n$ shell can be a candidate of CR generator 
at least up to $\sim 2\times 10^{18}$~eV.

\subsection{The case in shocks in circum-nuclear matter}

What about a further possibility of production of UHECRs
in such high-$n$ shells up to $\sim 10^{20}~{\rm eV}$
 in circum-nuclear matter which extends at larger scale?
Here, we briefly discuss it.

Schrwachter et al. (2013) reported the existence of 
the molecular hydrogen accretion flow in the inner 50~pc
of NGC1275 by the Gemini North telescope observation. 
The accretion flow is oriented perpendicular to the radio jet axis.
They interpret it as the outer part of a collisionally excited 
turbulent accretion flow with a number density of electron 
of $\sim 4\times 10^{3}~{\rm cm^{-3}}$.
If a strong shock drives this turbulent hydrogen accretion flow,
then a geometrically thick shell with its width $\delta R\sim 5~{\rm pc}$ is expected.
Then, such  high-$n$ shells could produce UHECRs with the energy of 
a few$\times 10^{19}~{\rm eV}$ together with the assumption
that the $B$-field strength is averaged by turbulence in the hydrogen accretion flow.
However, it seems natural to suppose that
$B$-field strength and  the shock propagation velocity 
may decrease at large scale.
Therefore, it is not clear whether 
the molecular hydrogen accretion flow observed by Schrwachter et al. (2013) 
is really a good site for producing UHECRs.


So far, we conservatively adopt the Berezhko's amplification factor (Berezhko 2008).
However, some previous work seems to indicate
higher amplification rate of the $B$-fields.
For example. Fraschetti (2013) examined 
the magnetic field amplification
driven by the motion of vortical eddies and 
the amplification factor of the field
can reach $\sim 10^{3}$ using reasonable parameters.
If this is the case in 3C~84, 
then there is a possibility of production of UHECRs
up to $\sim 10^{20}~{\rm eV}$.

\section{Summary}

In the present work,
we explored the physical properties of a fossil shell 
associated with fading radio lobe in 3C~84.
In our recent  work presented in I15,
we have modeled the dynamical and spectral 
evolution of fossil shells that  are identical to the forward shocks
propagating in the external medium
and found that 
the fossil shell emission overwhelms 
the fading radio lobe  after 
the injection of energy and fresh particles from 
the jet has swathed off.
In fact, the forward shock 
still continues to supply fresh electrons into the shell,
while the radio lobe rapidly fades away.
We apply this model to 3C~84.
Below we summarize the results.

\begin{itemize}

\item

The low-$n$ fossil shell made of
shocked ambient matter  with the  number density of $\sim 0.3~{\rm cm^{-3}}$
shows IC-dominated spectrum and 
it  can be compatible to the sensitivity of CTA.
Presumable TeV $\gamma$-ray emission from
the central compact region may compete with the fossil shell emission.
The brightness of TeV $\gamma$-ray depends on
the activity of the central compact region in which 
blazar region is included. Hence we need to choose a
low-activity phase of the blazar region for exploring the fossil shell emission
in TeV $\gamma$-ray band.
The predicted radio emission from this low-$n$ fossil shell is 
much below the typical sensitivity of VLBI.

\item

The high-$n$ shell made of
shocked torus with the density of $\sim 10^{3}~{\rm cm^{-3}}$
shows a brighter synchrotron spectrum in general
peaking at higher frequency and reaching  a higher luminosity 
than in the case of a  low-$n$ fossil shell.
In particular,
if magnetic field amplification is effective in the high-$n$ shell,
then the field strength conservatively reaches $\sim 17$~mG order.
Interestingly, the theoretical prediction of this fossil shell
and detection threshold at radio band is comparable in this case.
Hence, performing VLBI observations with higher dynamic range
is important for a first detection of fossil shell with high-$n$.

\item

We propose that 
the high-$n$ shell with $B$-fields 
amplification is a possible site for  UHECRs.
The predicted $E_{p,\rm max}$ in the 
high-$n$ shell in 3C~84 is about $2\times 10^{18}$~eV
when a  slow diffusion in the plasma torus takes place.
The value of $E_{p,\rm max}$ is proportional to 
the $B$-field strength, which is determined by
non-linear process of field amplification.
If the field amplification factor in 3C~84 is 
higher than that  derived by Berezhko (2008),
which is suggested by Fraschetti (2013),
then 3C~84 can be a possible site 
for UHECR  with the energy $\sim 10^{20}$~eV.

\item

An actual spectrum should be the sum of  high-$n$ and low-$n$ shells.
Therefore, cooperative observations between VLBI and CTA 
would be more effective and highly encouraged 
for exploring physical properties of the fossil shells in 3C~84
in great detail.

\end{itemize}


\bigskip
\leftline{\bf \large Acknowledgment}
\medskip

\noindent

We thank the anonymous referee for the review and suggestions
for improving the paper.
We sincerely thank M. Orienti for fruitful discussions and useful comments.
This research has made use of data from the MOJAVE database 
that is maintained by the MOJAVE team (Lister et al. 2009).
The VLBA
is operated by the US National Radio Astronomy Observatory
(NRAO), a facility of the National Science Foundation operated
under cooperative agreement by Associated Universities, Inc.
This work is partly based on observations made with the KaVA, 
which is operated by  the Korea Astronomy and Space Science
Institute (KASI) and the National Astronomical Observatory of Japan (NAOJ).
Data analysis was in part carried out on PC cluster 
 and computers at Center for Computational Astrophysics, NAOJ.
HI acknowledges the financial support of a Grant-in-Aid for Young Scientists (B:16K21630).
NK acknowledges the financial support of Grant-in-Aid for
Young Scientists (B:25800099).
HN is supported by MEXT KAKENHI Grant Number 15K17619.
Part of this work was done with
the contribution of the Italian Ministry of Foreign Affairs and
University and Research for the collaboration project between
Italy and Japan.



~\begin{table}
\centering
\caption{Physical quantities in 3C84}
\label{table:lobe}       
\scalebox{0.7}{
\begin{tabular}{lccc}
\hline\noalign{\smallskip}
{\bf quantities} & {\bf symbols} & {\bf values}&{\bf note}\\
\noalign{\smallskip}\hline\noalign{\smallskip}

 Eddington power&
 $L_{\rm Edd}$ &
 $ 1\times 10^{47}~{\rm ergs~s^{-1}}$&
 Scharwachter et al. (2013)
\\
 jet power&
 $L_{\rm j}$ &
 $ 5\times 10^{45}~{\rm ergs~s^{-1}}$&
 Heinz et al. (1998)
 \\
 age of dying radio lobe &
 $t$ &
 $ 50-60 ~{\rm years}$&
  in 2015, Nesterov et al. (1995)
 \\
 number density of ambient medium &
 $n_{\rm amb}$ &
 $0.3 ~{\rm cm^{-3}}$ & 
Fabian et al. (2006)
 \\
power-law index of ambient medium &
 $\alpha$ &
 $0$ & 
 Eq.~(\ref{eq:rho_ext}), flat
 \\
number density of plasma torus &
 $n_{\rm torus}$ &
 $1\times 10^{3}~{\rm cm^{-3}}$&
 O'Dea et al. (1984)
 \\
 radius (in 2015) &
 $R_{\parallel}$ &
 $9.3-16.6~{\rm pc}$ &  
Tavecchio et al. (2014); Fujita \& Nagai (2016)
\\
 propagation velocity &
 $v_{\parallel}$ &
 $0.3-0.5~c$  &
  Asada et al. (2006), Lister et al. (2013)
 \\
UV luminosity (accretion disk)&
 $L_{\rm UV}$ & 
$5\times 10^{42}~{\rm ergs~s^{-1}}$ &
KANATA
\\
IR luminosity (dust torus) &
$L_{\rm IR}$&
$ L_{\rm UV}/2$ &
Calderone et al. (2012)
 \\
Radio luminosity (central compact region)&
$L_{\rm 43G}$&
$ 4 \times 10^{42}~{\rm ergs~s^{-1}}$ &
Fig.~\ref{fig:KaVA43GHz}
 \\
Radio luminosity (central compact region)&
$L_{\rm 22G}$&
$ 2 \times 10^{42}~{\rm ergs~s^{-1}}$ &
VERA (Chida et al. 2015)
 \\
 Radio luminosity (fading radio robe)&
$L_{\rm lobe}$&
$ 1\times 10^{41}~{\rm ergs~s^{-1}}$ &
Fig.~\ref{fig:15G}
 \\
\noalign{\smallskip}\hline
\end{tabular}}
\end{table}

\begin{table}
\centering
\caption{Model Parameters}
\label{table:environment}       
\scalebox{0.7}{
\begin{tabular}{lcccc}
\hline\noalign{\smallskip}
{\bf quantities} & {\bf symbols} & {\bf low-$n$}&{\bf high-$n$ w/o $B$-amp}&{\bf high-$n$ w/ $B$-amp}\\
\noalign{\smallskip}\hline\noalign{\smallskip}

 number density of external matter&
 $n_{\rm ext}$&
 $n_{\rm ext}=n_{\rm amb}=0.3~{\rm cm^{-3}}$ &
  $n_{\rm ext}=n_{\rm torus}=1\times 10^{3}~{\rm cm^{-3}}$&
  $n_{\rm ext}=n_{\rm torus}=1\times 10^{3}~{\rm cm^{-3}}$
 \\
 magnetic field in the shell &
 $B_{\rm shell}$ &
 $0.1 {\rm mG}$&
 $0.1 {\rm mG}$&
 $17 {\rm mG}$
 \\
fraction of non-thermal electrons &
 $\epsilon_{\rm e,shell}$ &
 $0.1$&
 $0.1$&
 $0.01$
 \\
power-law index of injected electrons&
 $p_{\rm shell}$ &
 2&
 2&
 2
 \\
electron gyro-factor &
 $\xi_{\rm shell}$ &
 10&
 10&
 10
\\
volume filling factor in the shell&
 $f_{\rm amb}=f_{\rm torus}$ &
0.5&
0.5&
0.5
\\
\noalign{\smallskip}\hline
\end{tabular}}
\end{table}

\end{document}